# Giant barocaloric effects in natural rubber: A relevant step toward solid-state cooling

N. M. Bom[1], W. Imamura[1,2], E. O. Usuda[1,3], L. S. Paixão[1], A. M. G. Carvalho[1]*

[1]Laboratório Nacional de Luz Síncrotron (LNLS), Centro Nacional de Pesquisa em Energia e Materiais (CNPEM), CEP 13083-100, Campinas, SP, Brazil.

[2]Faculdade de Engenharia Mecânica, Universidade Estadual de Campinas (UNICAMP), CEP 13083-860, Campinas, SP, Brazil.

[3]Departamento de Ciências Exatas e da Terra, Universidade Federal de São Paulo, CEP 00972-270, Diadema, SP, Brazil.

**ABSTRACT:** Solid-state cooling based on *i*-caloric effects has shown to be a promising alternative to the conventional refrigeration devices. Only very recently, the research on barocaloric materials is receiving a deal of attention due to the demonstration of giant barocaloric effects in shape-memory alloys. Regarding polymers, there is still a lack of literature, despite their high caloric potential. Thus, we present here giant barocaloric effects in natural rubber, a low-cost and environmental friendly elastomer polymer. The maximum values of entropy and temperature changes are larger than those previously reported for any promising barocaloric material. Moreover, the huge normalized temperature change and refrigerant capacity exhibited by natural rubber confirm its high potential for cooling applications. We also verify a relevant dependence of the barocaloric effect on the glass transition in natural rubber. Our findings suggest that commercial refrigeration devices based on barocaloric effects from elastomer polymers can be envisaged in the near future.

Natural rubber (also known as "India rubber") is a polymer formed by continuous chains of the organic compound isoprene ($C_5H_8$). It presents unique physical and chemical properties, the most important being the elasticity: natural rubber (NR) can be reversibly stretched several times its initial length upon relatively low stresses. Although Mesoamericans already used stabilized forms of NR since antiquity,[1] it was only after the discovery of sulfur vulcanization, in the nineteenth century,[2] that NR has widely spread into industrial applications and commercial products. The explanation is that rubber-like elasticity is only exhibited if a certain number of cross-linkages are present between the polymer chains at some points,[3] which are introduced by the vulcanization process. Polymer materials presenting a mechanical behavior similar to NR (rubber-like properties), independently of their chemical constitution, are called elastomers.

In addition to its remarkable mechanical properties, NR presents another peculiarity: it warms up when rapidly stretched, and cools down when the stress is released. This effect was first observed by John Gough in 1805,[4] and further studied by Joule[5] for NR and other classes of materials. In fact, both scientists described what it is now designated as elastocaloric effect ($\sigma_e$-CE), the first $i$-caloric effect ever reported. The $i$-caloric effects refer to the isothermal entropy change ($\Delta S_T$) or the adiabatic temperature change ($\Delta T_S$) registered in a material upon the application/removal of an external field ("$i$" stands for intensive thermodynamic variables). Depending on the nature of the field, the $i$-caloric effects can be categorized as magnetocaloric effect ($h$-CE), electrocaloric effect ($e$-CE) and barocaloric effect ($\sigma_b$-CE), besides the aforementioned $\sigma_e$-CE. Both $\sigma_e$-CE and $\sigma_b$-CE are collectively described as mechanocaloric effects ($\sigma$-CE), which are driven by mechanical stresses.[6–8] Solid-state cooling based on $i$-caloric effects appeared as a promising alternative for the conventional vapor-compression devices, matching the current energy-saving and eco-friendliness requirements for technological development.[7,9–12]

In this context, $\sigma$-CE is receiving a great deal of attention in the last years as consequence of the work on the other caloric effects[10] – extensively studied since the observation of the giants $h$-CE[13] and $e$-CE[14]. Several shape-memory alloys (SMAs) exhibit giant values of $\sigma$-CE around room temperature.[15–19] Regarding barocaloric materials, promising properties have been recently reported: $\Delta T$ = 8.4 K for $MnCoGe_{0.99}In_{0.01}$ ($\Delta\sigma$ = 300 MPa);[20] $\Delta T$ = 16 K and $\Delta S_T$ = -74 J $kg^{-1}$ $K^{-1}$ for $(MnNiSi)_{0.62}(FeCoGe)_{0.38}$ ($\Delta\sigma$ = 270 MPa).[21] Below room temperature, large $\Delta S_T$ values are observed for $(NH_4)_2SO_4$ ($\Delta S_T$ = -60 J $kg^{-1}$ $K^{-1}$ for $\Delta\sigma$ = 100 MPa).[22] The good performance of SMAs stimulated, for example, the development of a sophisticated scientific setup for elastocaloric investigation by Schimdt et al.[23] and a regenerative elastocaloric heat pump prototype by Tusek et al.[24] Alternatively to SMAs, two families of polymers show promising mechanocaloric potential: elastomers and PVDF-based polymers.[12] Among elastomers, NR is prominent due to its non-toxicity, sustainability and low-cost, also presenting long fatigue life.[25] Moreover, large temperature changes ($\Delta T \sim 12$ K) on fast stretching of NR were already observed in the 1940 decade by Dart et al.,[26] evidencing its high caloric potential. Despite these favorable characteristics, the research on NR in view of cooling applications is still an incipient topic. Nevertheless, significant advances on the elastocaloric investigation of NR have been made in recent years.[27–29]

On the other hand, $\sigma_b$-CE is the least studied $i$-caloric effect, with a lack of literature for many classes of materials (such as polymers). Concerning PVDF polymers, there a single work reporting experimental results for PVDF-TrFE-CTFE ($\Delta T \sim 18$ K and $\Delta S_T \sim 120$ J kg$^{-1}$ K$^{-1}$ for $\Delta \sigma = 200$ MPa), obtained from indirect measurements.[30] In the specific case of NR, only very recently experimental results of $\sigma_b$-CE were reported.[31,32] The entropy/temperature changes associated to $\sigma_b$-CE are induced by the application of isotropic mechanical stress (isostatic pressure). Based on this principle, Figure 1 schematically illustrates the barocaloric cooling process.

In the present study, we systematically investigate the $\sigma_b$-CE in NR around room temperature at low applied pressures, in order to explore the encouraging perspectives for this material concerning future solid-state cooling technology. The experiments were carried out in a customized setup developed by our group.[31] Giant $\sigma_b$-CE were observed in NR, exceeding all previously reported values. We compared NR with other barocaloric materials in literature by the calculation of caloric performance parameters, confirming the high barocaloric potential of NR. Moreover, our results show the influence of the glass transition on the adiabatic temperature change in NR.

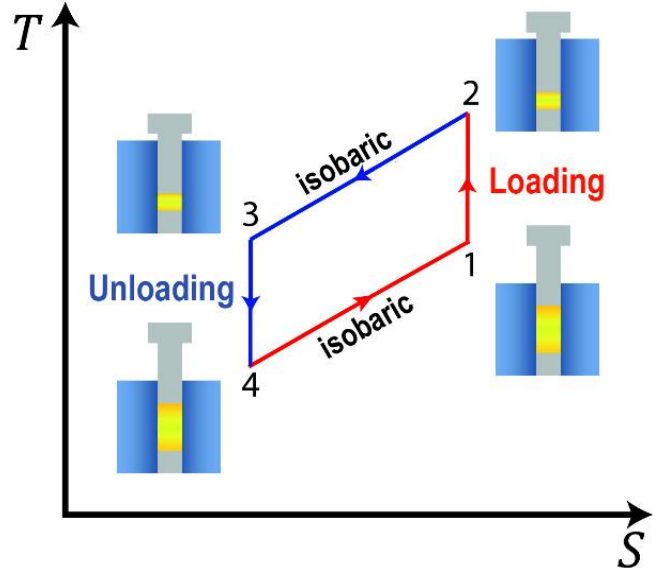

**Figure 1:** Schematic illustration of the barocaloric cooling process, following the Brayton cycle. The yellow rectangles represent the solid refrigerant during the cycles. At the first step (1 → 2), the material is adiabatically compressed, increasing the sample temperature. Then, the heat gradually flows out from the sample to the heat sink and the temperature decreases down to initial temperature, as the applied pressure is kept constant (2 → 3). In the third step (3 → 4), the sample is cooled down as result of the adiabatic release of the load. Finally, (4 → 1), the heat is absorbed from the surroundings and the sample returns to its initial state.

$\Delta T_S$ of NR in the decompression process, as function of the initial temperature, was measured within the 43.4(9)–390(12) MPa pressure range (Fig. 2a). For the 26-87 MPa range, $\Delta T_S$ remains stable up to 263 K, slight increasing above room temperature. A similar pattern is observed for 173 MPa, but with a sharp decrease in $\Delta T_S$ at 223 K. On the other hand, for the two highest pressures (273 and 390 MPa) $\Delta T_S$ curves present a strong dependence with initial temperature, reaching the maximum of ~25 K (390 MPa, 314 K). This value surpasses the barocaloric $\Delta T$ obtained for PVDF-TrFE-CTFE[30] or for any other intermetallic barocaloric materials reported so far.[8]

In the three highest pressures, we can observe a temperature threshold, below which $\Delta T_S$ values sharply reduce (Fig. 2a). That threshold shifts towards higher temperatures as pressure increases. Analyzing the glass transition ($T_g$) as a function of pressure (Fig. 2b), we verify that $T_g$ also shifts to higher temperatures as pressure increases at a rate of $dT_g/\sigma_b$ = 0.16(2) K MPa$^{-1}$. Moreover, $T_g$ values are in accordance with the temperature thresholds presented by the curves in Fig. 2a. The deleterious influence of $T_g$ on $\Delta T_S$ can be explained by the fact that, at $T_g$ or below, the derivative $(\partial\varepsilon/\partial T)_\sigma$ is mightily reduced.

Thus, it also should affect the isothermal entropy change values in the same temperature range.

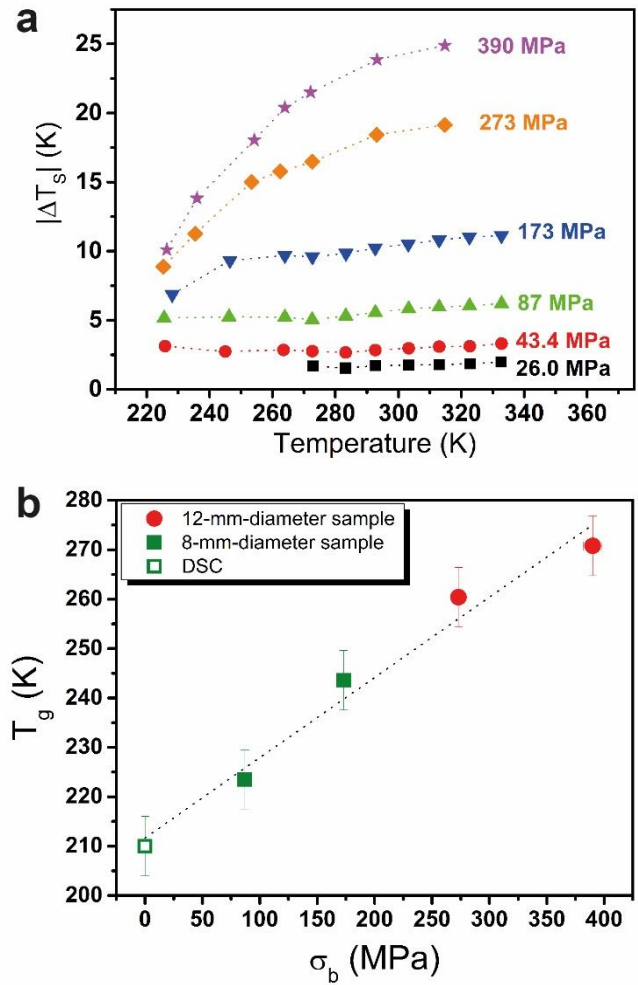

**Figure 2:** a) Adiabatic temperature change ($\Delta T_S$) vs. initial temperature. Values measured during the decompression process of NR at different pressure variations (26.0(5), 43.4(9), 87(2), 173(3), 273(8) and 390(12) MPa); the dotted lines are guides for the eyes. b) Glass transition temperature ($T_g$) vs. applied stress ($\sigma_b$); opened symbol is $T_g$ at atmospheric pressure, from DSC measurement, and closed symbols are $T_g$ at different pressures, from ε vs T data, on heating process; squares are for 8-mm-diameter sample, and circles are for 12-mm-diameter sample; the dotted line corresponds to the linear fitting.

The direct measurement of entropy change ($\Delta S$) usually represents a problem for caloric experiments: concerning $\sigma_b$-CE, only quasi-direct measurements of $\Delta S$ were reported up to now.[8] Nevertheless, this issue can be partially addressed by an indirect quantification of the caloric effects. This method is implemented by measuring the derivative $(\partial\varepsilon/\partial T)_\sigma$ and using a Maxwell's relation to obtain the following expression for the entropy change[33,34]:

$$\Delta S_T(T, \Delta\sigma) = -\frac{1}{\rho_0} \int_{\sigma_1}^{\sigma_2} \left(\frac{\partial\varepsilon}{\partial T}\right)_\sigma d\sigma \quad (1)$$

where $\sigma$ and $\rho_0$ are the compressive stress and the density of the sample at atmospheric pressure, respectively. $\varepsilon$ is the compressive strain, defined as $\varepsilon \equiv \Delta l/l_0$ (where $\Delta l$ is length change of the sample and $l_0$ is the initial length). $(\partial\varepsilon/\partial T)_\sigma$ was determined by the $\varepsilon$ vs. T curves measured at constant pressures (Fig. 2, *Supplementary Information*).

Giant $\Delta S_T$ values were obtained at relatively low pressures on cooling, as shown in Fig. 3. At 290 K, the values calculated from eq. 1 are $\Delta S_T = -50(7)$ J kg$^{-1}$ K$^{-1}$ for $\Delta\sigma = 87(2)$ MPa and supergiant $\Delta S_T = -96(11)$ J kg$^{-1}$ K$^{-1}$ for $\Delta\sigma = 173(3)$ MPa, leading to the normalized entropy changes of 0.58(8) and 0.56(6) kJ kg$^{-1}$ K$^{-1}$ GPa$^{-1}$, respectively.. It is worth noticing that $\Delta S_T$ increases as temperature decreases, showing a clear trend of increasing at temperatures below 290 K, mainly for $\Delta\sigma = 173$ MPa.

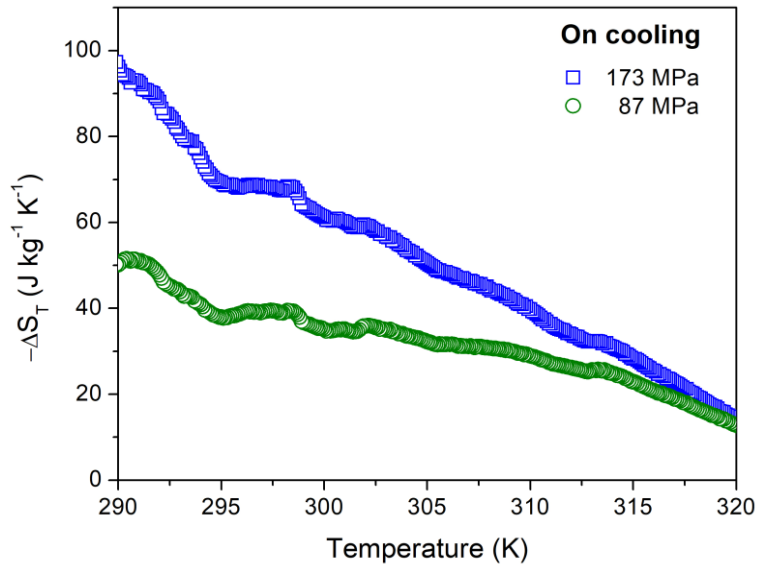

**Figure 3:** Isothermal entropy change ($\Delta S_T$) as a function of temperature. Values for $\Delta\sigma = 87(2)$ and 173(3) MPa, obtained from $\varepsilon$ vs. T data for NR at constant pressures (Supplementary Information, Fig. 2). The estimated errors are (1 J kg$^{-1}$·K$^{-1}$ + 12%) and (1 J·kg$^{-1}$ K$^{-1}$ + 10%) for $\Delta\sigma = 87$ and 173 MPa, respectively. The error bars are omitted for the sake of clarity.

Usually, the magnitude of entropy (S) and temperature (T) changes in *i*-caloric materials are proportional to the intensive thermodynamic variables (i.e., the maximum applied fields), following a power law in the form of $\Delta X(T,i) = a_X i^{n_X}$ (X: S or T). Examples of this behavior are reported by Oesterreicher and Parker[35] for *h*-CE ($\Delta S \propto$ $H^{2/3}$), and by Lu et al.[36] for $e$-CE ($\Delta T \propto E^2$ at low electric fields, $\Delta T \propto E^{2/3}$ for higher fields). Based on the same concept, we propose the following power law for the $\sigma_b$-CE:

$$-\Delta X(T, \sigma_{max}) = a_X \sigma_{max}^{n_X} \quad (2)$$

where $a_X$ is the constant of proportionality, $\sigma_{max}$ is the maximum value of the applied/released pressure and $n_X$ is the power law exponent. Equation (2) is a generalization of the expression previously used by Usuda et al.[32] to fit the barocaloric $\Delta T_S$ as a function of $\sigma_{max}$ in NR.

Another approach to take into account the relationship between the compressive stress and the entropy change comes from a thermodynamic model based on Landau's theory of elasticity.[37] Let us consider the power series expansion for the Helmholtz free energy:

$$F(T, \varepsilon_{ij}) = F_0(T) - B\alpha(T - T_0)\varepsilon_{kk} - \frac{1}{2}B[\beta(T - T_0)^2 - 1]\varepsilon_{kk}^2$$
$$+G\left(\varepsilon_{ij} - \frac{1}{3}\varepsilon_{kk}\delta_{ij}\right)^2 \quad (3)$$

where $F_0$ is the free energy of the unstrained body, $\alpha$ is the thermal expansion coefficient, $B$ and $G$ are the bulk and shear moduli, respectively. The stress and entropy are obtained from the free energy above through its derivative with respect to strain and temperature, respectively. The boundary condition for confined compression states that only one diagonal component of the strain tensor (e.g. $\varepsilon_{zz}$) is non-zero. Therefore, the entropy change can be written as (details in the *Supplementary Information*):

$$\Delta S(T, \sigma) = a_1(T)\sigma + a_2(T)\sigma^2 \quad (4)$$

The experimental -$\Delta T_S$ vs. $\sigma_{max}$ curve for NR at 293 K and the correspondent fitted model (equation (2)) are displayed in Fig. 4a, with $a_T = 61(5)$ K GPa$^{-n_T}$ and $n_T = 0.98(6)$. The values obtained from the curve fitting at other temperatures are listed in the *Supplementary Information* (Table 1). Experimental data for $\Delta S_T$ as a function of $\sigma_{max}$ is shown in Fig. 4b. The curves were fitted both by the power law (equation (2)) and the quadratic function (equation (4)) derived from Helmholtz free energy (equation (3)). The fitting parameters obtained from this calculation are listed in Table 2 (*Supplementary Information*). Both fittings are in very good agreement with the experimental data at this specific temperature range.

We could expect that the values of the fitting parameters $n_T$ and $n_S$ obtained from $-\Delta T_S$ vs $\sigma_{max}$ and $-\Delta S_T$ vs $\sigma_{max}$ curves, respectively, are very close, but they are not. This fact can be explained by the distinct experimental procedures used to obtain the $\Delta T_S$ and $\Delta S_T$ curves, which imply in essentially different thermodynamic processes: quasi-adiabatic and isobaric processes, respectively. The thermodynamic irreversibilities associated to each isobaric ε vs. T curve may result in $\Delta S_T$ values different from those that would be obtained from reversible processes. An example on how different experimental protocols (or thermodynamic processes) can affect the $\Delta S_T$ values of *i*-caloric effects was reported by Carvalho and co-authors[38].

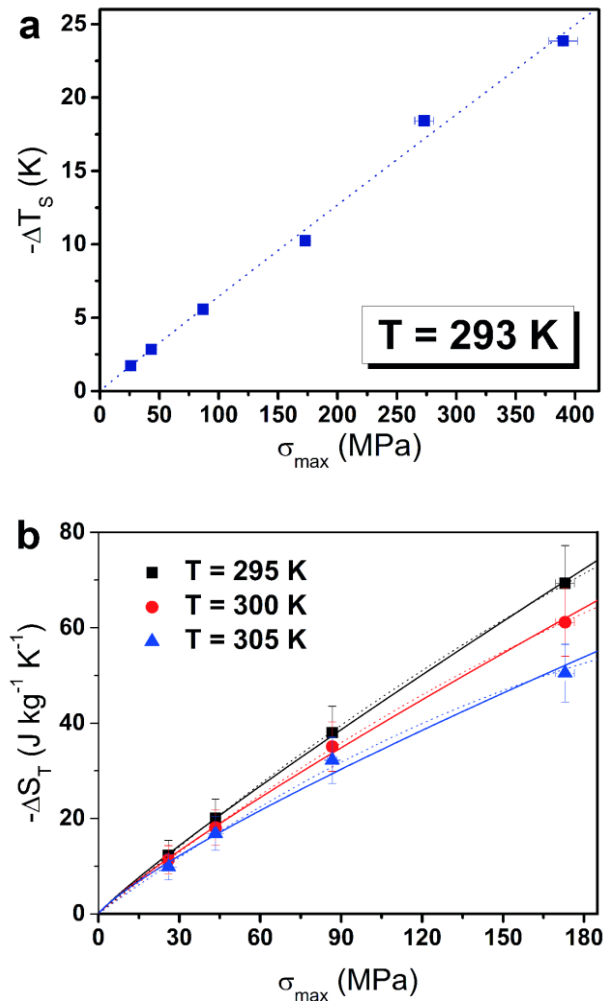

**Figure 4:** $\sigma_b$-CE as a function of maximum applied pressure ($\sigma_{max}$). (a) $\Delta T_S$ (decompression) vs. released pressure for initial temperature of 293 K; the dotted line corresponds to the fitting using equation (2); the pressure error is estimated in 2% for pressures up to 173 MPa and 3% above 173 MPa. (b) $\Delta S_T$ vs. maximum applied pressure within the 305–295 K temperature range; the dotted and solid lines correspond to the fittings using equation (2) and equation (4), respectively.

The physicochemical mechanisms underlying the observed $\sigma_b$-CE (i.e., pressure-induced entropy or temperature changes) in NR are likely related to the rearrangement of polymer chains during the compression/decompression process. An analogy can be drawn with $\sigma_e$-CE experiments, where the entropy decreases due the alignment of polymer chains induced by stretching, leading to the increase of the temperature in NR.[27] In the case of barocaloric experiments, it is plausible to suppose the confined compression of NR leads to a decrease in its free volume (total volume minus volume occupied by chains), constraining the degrees of freedom of the polymer chains, particularly the rotation of the chemical bonds and the movements of the chains backbones. As consequence, the entropy and temperatures variations associated to the $\sigma_b$-CE are observed even in the absence of phase transitions, on the contrary of what was suggested elsewhere.[12] The proposed qualitative model also clarifies the strong link between $T_g$ and $\Delta T_S$, as evidenced in Fig.2. $T_g$ marks the transition point from a rubbery state to a rigid state in a polymer. From a molecular point of view, it means a sharp decrease in motion of large segments of polymer chains when temperature reaches $T_g$ (on cooling), explaining why $\Delta T_S$ is significantly lower below this temperature.

The present results concerning absolute values of $\Delta T_S$ and $\Delta S_T$ are quite impressive when compared with those previously reported. Despite this, other relevant features should be considered for a proper evaluation of a specific material in view of cooling applications. For instance, the requirement of large applied pressures can represent a practical limitation for commercial devices, even if the material exhibits large $\Delta T_S$. In order to take both parameters into account, we display in Fig. 5a the normalized temperature change ($|\Delta T_S/\Delta\sigma|$) as a function of the initial temperature of NR (263–322 K temperature range); and as function of measured $|\Delta T_S|$ values corresponding to those temperatures (Fig. 5b). These results are compared with published data for some prominent barocaloric materials. The $|\Delta T_S/\Delta\sigma|$ in NR reaches the maximum value of ~63 K GPa$^{-1}$ at 322 K and $|\Delta T_S|$~11 K. It is noteworthy that all values for NR are significantly higher than those reported for any other barocaloric material.

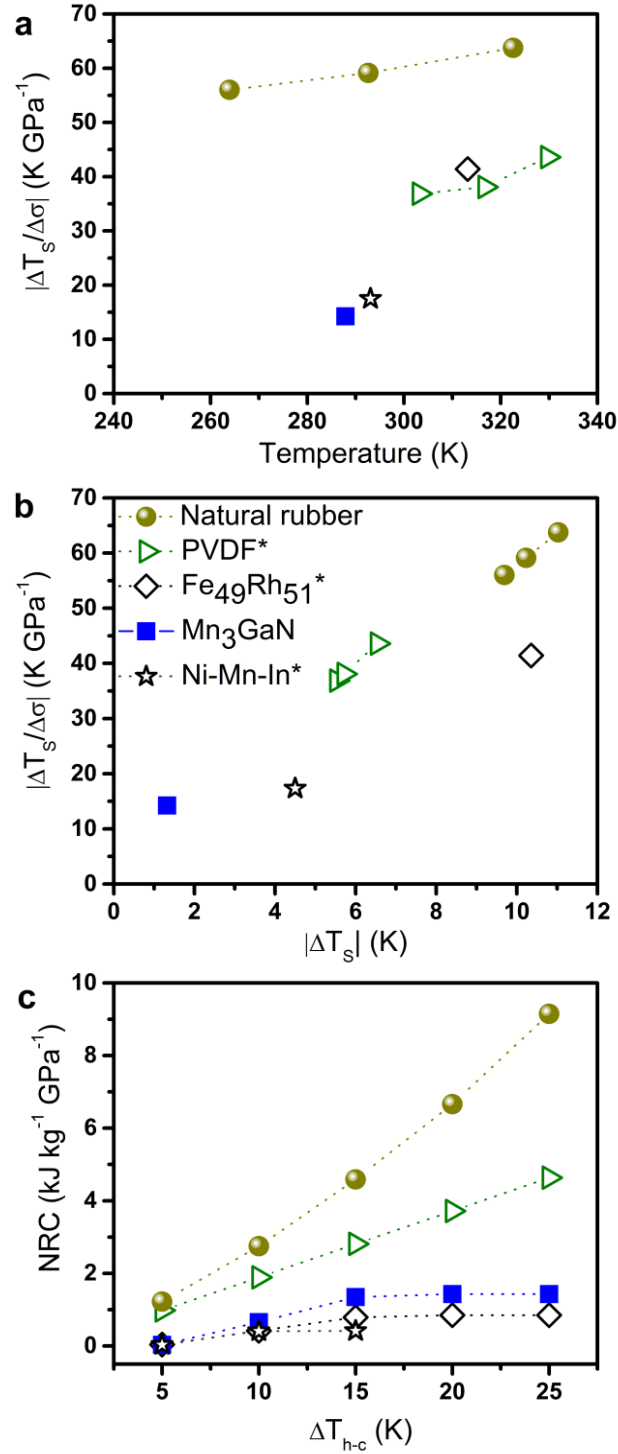

**Figure 5:** Barocaloric properties for materials with large or giant $\sigma_b$-CE around room temperature (250–330 K). Normalized temperature change ($|\Delta T_S/\Delta\sigma|$) as function of (a) initial temperature and (b) measured $|\Delta T_S|$ values. The materials and applied pressures are: NR ($|\Delta\sigma| =$ 173 MPa); PVDF[30] ($|\Delta\sigma| =$ 150 MPa); $Mn_3GaN$[39] (maximum reported for $|\Delta\sigma| =$ 93 MPa); Ni-Mn-In[16] (maximum reported for $|\Delta\sigma| =$ 260 MPa); $Fe_{49}Rh_{51}$[40] (maximum reported for $|\Delta\sigma| =$ 250 MPa). *Indirect determination. (c) Normalized refrigerant capacity (NRC $\equiv |RC/\Delta\sigma|$) as a function of $\Delta T_{h-c} \equiv T_{hot} - T_{cold}$ (temperature difference between hot reservoir and cold reservoir): NR ($T_{hot} =$ 315 K, $|\Delta\sigma| =$ 130 MPa); PVDF ($T_{hot} =$ 330 K, $|\Delta\sigma| =$ 150 MPa); $Mn_3GaN$ ($T_{hot} =$ 295 K, $|\Delta\sigma| =$ 139 MPa); Ni-Mn-In ($T_{hot} =$ 300 K, $|\Delta\sigma| =$ 160 MPa); $Fe_{49}Rh_{51}$ ($T_{hot} =$ 325 K, $|\Delta\sigma| =$ 160 MPa).

Aiming to compare the barocaloric materials considering the cooling power for different temperature spans, we also calculated the normalized refrigerant capacity (NRC) as a function of the temperature difference between hot reservoir and cold reservoir ($\Delta T_{h-c} \equiv T_{hot} - T_{cold}$), using the following relationship:

$$NRC(\Delta T_{h-c}, \Delta\sigma) = \left|\frac{1}{\Delta\sigma}\int_{T_{cold}}^{T_{hot}}\Delta S_T(T,\Delta\sigma)dT\right| \quad (5)$$

For NR, the hot reservoir was fixed at 315 K. In Fig. 5c, it is possible to observe that NRC at $\Delta T_{h-c} = 5$ K for NR is approximately the same as for PVDF and higher than the other materials displayed in the graph. However, for higher $\Delta T_{h-c}$ values the NRC for NR increases, reaching the striking value of ~9 kJ kg$^{-1}$ GPa$^{-1}$ at $\Delta T_{h-c} = 25$ K. Besides, the NRC curve for NR clearly indicates a tendency of increase for higher temperature spans, contrasting with the behavior of other compared materials.

In summary, our results demonstrate that the barocaloric potential of NR is higher than any prominent barocaloric material investigated up to now. Very large pressure-induced changes in the temperature and entropy were obtained at relatively low pressures, reaching the giant $\Delta T_S$ value of ~ 25 K (for $|\Delta\sigma|$ of 390 MPa, at 314 K) and supergiant $\Delta S_T$ value of 96 J kg$^{-1}$ K$^{-1}$ (for $|\Delta\sigma|$ of 173 MPa, at 290 K). These barocaloric effects were observed in the absence of phase transitions during the compression/decompression cycles, on the contrary of what is verified for intermetallic compounds and semicrystalline polymers. A strong dependence of the barocaloric effect on the glass transition in natural rubber is verified, explaining the sharp decrease in $\Delta T_S$ for lower temperatures. Also, we propose a thermodynamic relationship between the compressive stress and the entropy change, based on the Landau's theory of elasticity. Finally, the normalized barocaloric parameters are also huge: the maximum observed normalized temperature change is $|\Delta T_S/\Delta\sigma| \approx 63$ K GPa$^{-1}$ (at 322 K) and the normalized refrigerant capacity NRC $\approx$ 9 kJ kg$^{-1}$ GPa$^{-1}$ (for $\Delta T_{h-c} = 25$ K). All these values exceed those previously published results regarding barocaloric effect obtained from direct measurements, for any class of materials or experimental conditions considered. It is expected that the present findings could inspire further developments related to barocaloric materials and prototypes, boosting the progress of solid-state cooling technology soon.

## SUPPORTING INFORMATION

Experimental details and preparation methods, FTIR data, experimental strain vs. temperature curves, fitting parameters data, derivation of the ΔS expression and demonstration of the isostatic condition for NR compression (PDF).

## AUTHOR INFORMATION

**Corresponding author:** A.M.G.C., E-mail: alexandre.carvalho@lnls.br

**Author contributions:** A.M.G.C. conceived the study and led the project. N.M.B, A.M.G.C and L.S.P. wrote the paper. W.I. and E.O.U. executed the experiments. N.M.B., A.M.G.C., W.I. and E.O.U. planned the experiments. All authors discussed the results and analyzed the data.

**Competing financial interests:** The authors declare no competing financial interests.

**Data availability:** The datasets generated during the current study are available from the corresponding author on reasonable request.

## ACKNOWLEDGEMENTS

The authors acknowledge financial support from FAPESP (project number 2012/03480-0), CNPq, CAPES, LNLS and CNPEM. The authors also thank Prof. Angelo M. S. Gomes for DSC measurements.

## REFERENCES

(1) Hosler, D.; Burkett, S. L.; Tarkanian, M. J. *Science.* **1999**, *284* (5422), 1988–1991.

(2) Ghosh, P.; Katare, S.; Patkar, P.; Caruthers, J. M.; Venkatasubramanian, V.; Walker, K. A. *Rubber Chem. Technol.* **2003**, *76* (3), 592–693.

(3) Treloar, L. R. G. *The physics of rubber elasticity*, 3rd ed..; Oxford University Press: London, 1975.

(4) Gough, J. A. *Mem. Lit. Phyiosophical Soc. Manchester* **1805**, *1*, 288–295.

(5) Joule, J. P. *Phil. Trans. R. Soc. Lond.* **1859**, *149* , 91–131.

(6) Lu, B.; Liu, J. *Sci. Bull.* **2015**, *60* (17), 1638.

(7) Crossley, S.; Mathur, N. D.; Moya, X. *AIP Adv.* **2015**, *5*, 67153.

(8) Mañosa, L.; Planes, A. *Adv. Mater.* **2017**, 1603607.

(9) Mañosa, L.; Planes, A.; Acet, M. *J. Mater. Chem. A* **2013**, *1*, 4925–4936.

(10) Moya, X.; Kar-Narayan, S.; Mathur, N. D. *Nat. Mater.* **2014**, *13*, 439–450.

(11) Takeuchi, I.; Sandeman, K. *Phys. Today* **2015**, *68* (12), 48–54.

(12) Xie, Z.; Sebald, G.; Guyomar, D. **2016**, arXiv:1604.04479.

(13) Pecharsky, V. K.; Gschneidner, Jr., K. A. *Phys. Rev. Lett.* **1997**, *78* (23), 4494–4497.

(14) Mischenko, A. S.; Zhang, K.; Scott, J. F.; Whatmore, R. W.; Mathur, N. D. *Science (80-. ).* **2006**, *311*, 1270–1271.

(15) Bonnot, E.; Romero, R.; Mañosa, L.; Vives, E.; Planes, A. *Phys. Rev. Lett.* **2008**, *100*, 125901.

(16) Mañosa, L.; González-alonso, D.; Planes, A.; Bonnot, E.; Barrio, M.; Tamarit, J.; Aksoy, S.; Acet, M. *Nat. Mater.* **2010**, *9*, 478–481.

(17) Millán-Solsona, R.; Stern-Taulats, E.; Vives, E.; Planes, A.; Sharma, J.; Nayak, A. K.; Suresh, K. G.; Mañosa, L. *Appl. Phys. Lett.* **2014**, *105*, 241901.

(18) Stern-Taulats, E.; Planes, A.; Lloveras, P.; Barrio, M.; Tamarit, J. L.; Pramanick, S.; Majumdar, S.; Frontera, C.; Mañosa, L. *Phys. Rev. B - Condens. Matter Mater. Phys.* **2014**, *89*, 214105.

(19) Yuce, S.; Barrio, M.; Emre, B.; Stern-Taulats, E.; Planes, A.; Tamarit, J. L.; Mudryk, Y.; Gschneidner, K. A.; Pecharsky, V. K.; Mañosa, L. *Appl. Phys. Lett.* **2012**, *101* (7).

(20) Wu, R.-R.; Bao, L.-F.; Hu, F.-X.; Wu, H.; Huang, Q.-Z.; Wang, J.; Dong, X.-L.; Li, G.-N.; Sun, J.-R.; Shen, F.-R.; Zhao, T.-Y.; Zheng, X.-Q.; Wang, L.-C.; Liu, Y.; Zuo, W.-L.; Zhao, Y.-Y.; Zhang, M.; Wang, X.-C.; Jin, C.-Q.; Rao, G.-H.; Han, X.-F.; Shen, B.-G. *Sci. Rep.* **2015**, *5*, 18027.

(21) Samanta, T.; Lloveras, P.; Saleheen, A. U.; Lepkowski, D. L.; Kramer, E.; Dubenko, I.; Adams, P. W.; Young, D. P.; Barrio, M.; Tamarit, J.-L.; Ali, N.; Stadler, S. **2016**, arXiv:1011.1669v3.

(22) Lloveras, P.; Stern-Taulats, E.; Barrio, M.; Tamarit, J.-L.; Crossley, S.; Li, W.; Pomjakushin, V.; Planes, A.; Mañosa, L.; Mathur, N. D.; Moya, X. *Nat. Commun.* **2015**, *6*, 8801.

(23) Schmidt, M.; Schu, A.; Seelecke, S. *Int. J. Refrig.* **2015**, *54*, 88–97.

(24) Tušek, J.; Engelbrecht, K.; Eriksen, D.; Dall'Olio, S.; Tušek, J.; Pryds, N. *Nat. Energy* **2016**, *1* (10), 16134.

(25) Sebald, G.; Xie, Z.; Guyomar, D. *Philos. Trans. R. Soc. London A Math. Phys. Eng. Sci.* **2016**, *374* (2074), 439–450.

(26) Dart, S. L.; Anthony, R. L.; Guth, E. *Ind. Eng. Chem.* **1942**, *34* (11), 1340–1342.

(27) Guyomar, D.; Li, Y.; Sebald, G.; Cottinet, P.; Ducharne, B.; Capsal, J. *Appl. Therm. Eng.* **2013**, *57*, 33–38.

(28) Xie, Z.; Sebald, G.; Guyomar, D.; Xie, Z.; Sebald, G.; Guyomar, D. *Appl. Phys. Lett.* **2015**, *107*, 81905.

(29) Xie, Z.; Sebald, G.; Guyomar, D. *Appl. Phys. Lett.* **2016**, *108*, 41901.

(30) Patel, S.; Chauhan, A.; Vaish, R.; Thomas, P. *Appl. Phys. Lett.* **2016**, *108*, 72903.

(31) Bom, N. M.; Usuda, E. O.; Guimarães, G. M.; Coelho, A. A.; Carvalho, A. M. G. *Rev. Sci. Instrum.* **2017**, *88* (4), 46103.

(32) Usuda, E. O.; Bom, N. M.; Carvalho, A. M. G. *Eur. Polym. J.* **2017**, *92*, 287–293.

(33) Liu, Y.; Infante, I. C.; Lou, X.; Bellaiche, L.; Scott, J. F.; Dkhil, B. *Adv. Mater.* **2014**, *26* (35), 6132–6137.

(34) Vopson, M. M. *J. Phys. D. Appl. Phys.* **2013**, *46* (34), 345304.

(35) Oesterreicher, H.; Parker, F. T. *J. Appl. Phys.* **1984**, *55* (12), 4334–4338.

(36) Lu, S. G.; Rozic, B.; Zhang, Q. M.; Kutnjak, Z.; Pirc, R. *Appl. Phys. A Mater. Sci. Process.* **2012**, *107* (3), 559–566.

(37) Landau, L. D.; Lifshitz, E. M. *Theory of Elasticity: Vol. 7 of Course of Theoretical Physics*, 2nd ed.; Pergamon Press: Oxford, 1970.

(38) Carvalho, A. M. G.; Coelho, A. A.; Gama, S.; von Ranke, P. J.; Alves, C. S. *J. Appl. Phys.* **2008**, *104*, 63915.

(39) Matsunami, D.; Fujita, A.; Takenaka, K.; Kano, M. *Nat. Mater.* **2015**, *14*, 73–78.

(40) Stern-Taulats, E.; Gràcia-Condal, A.; Planes, A.; Lloveras, P.; Barrio, M.; Tamarit, J. L.; Pramanick, S.; Majumdar, S.; Mañosa, L. *Appl. Phys. Lett.* **2015**, *107*, 152409.

# Supporting Information for

Giant Barocaloric Effects in Natural Rubber: A Relevant Step toward Solid-State Cooling

N. M. Bom, W. Imamura, E. O. Usuda, L. S. Paixão, A. M. G. Carvalho

correspondence to: alexandre.carvalho@lnls.br

**This PDF file includes:**

Materials and Methods
Figures S1, S2 and S3
Tables S1 and S2
Appendix A
Appendix B
References

**Materials and Methods**

Samples

The natural rubber (NR) samples were prepared from a pre-vulcanized latex resin (from Siquiplas), cast into a cylindrical plaster mold. A resin feeder was used to prevent the formation of cavities due to the shrinkage of the latex while drying. We made two samples with the following dimensions: 12 mm (diameter) and 19.5 mm (length); 8 mm (diameter) and 21.1 mm (length). The density of the samples is 902(7) kg m$^{-3}$. For pressures $\leq$ 173 MPa, we used the 12 mm-diameter sample; above 173 MPa, the 8-mm-diameter sample was used. We characterized the 12-mm-diameter sample via Fourier transform infrared spectroscopy (FTIR) from 450 to 4000 cm$^{-1}$, with a fixed step of 2 cm$^{-1}$, using a FTIR spectrometer from PerkinElmer® (model Spectrum Two). The absorption bands observed in the spectrum (Fig. S1) are typical of NR samples.

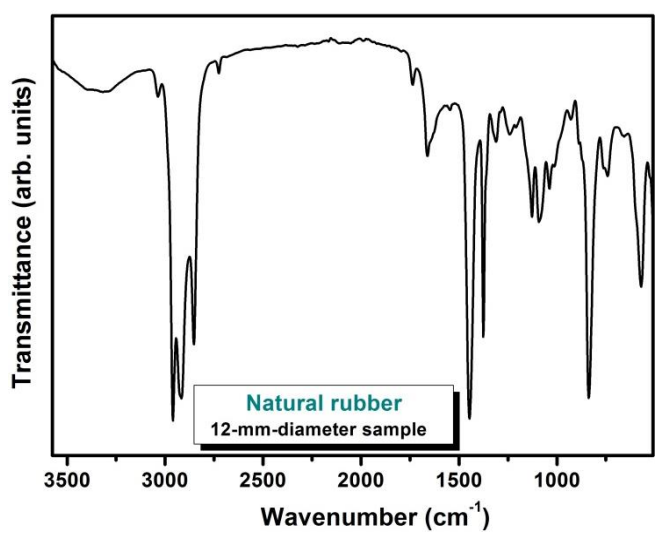

Figure S1. FTIR spectrum of the 12-mm-diameter sample of natural rubber.

Experimental setup

The experimental setup consists of a customized piston-cylinder carbon-steel chamber surrounded by a copper coil, enabling the circulation of cooling/heating fluids (water or liquid nitrogen, for instance). Two tubular heating elements (NP 38899, HG Resistências), placed in the proper holes in the chamber, are responsible for thermal stability when liquid nitrogen is used. Temperature is measured by two type-K thermocouples. A thermostatic bath (TE 184, Tecnal) was used to pump water in the copper coil above 280 K. Below 280 K, liquid nitrogen was employed to cool down the sample. Uniaxial load is applied by a manual 15,000-kgf hydraulic press (P15500, Bonevau). A load cell (3101C, ALFA Instrumentos) measures the contact force. Sample displacement is probed by a precise linear length gauge (METRO 2500, Heidenhain Co). Temperatures are collected and controlled (if heating elements are used) by Cryogenic Temperature Controller (Model 335, Lake Shore Cryotronic). This system is described in details by Bom et al (ref. 31, main text).

Description of the barocaloric experiments

The direct measurements of barocaloric temperature changes ($\Delta T_S$) were obtained by the following procedure: i) the sample was submitted to compressive stresses quasi-adiabatically, resulting in an immediate increase in temperature; ii) the load was kept constant, until the temperature decreases down to the initial temperature; iii) the load was released adiabatically, causing an abrupt decrease in the sample's temperature. $\Delta T_S$ curves for 273 and 390 MPa were measured from the maximum temperature (~314 K) down to minimum temperature (~223 K). Before starting the actual measurements, we have always performed several cycles in the maximum pressure until stabilizing the $\Delta T_S$ value. The experiments were carried out only when the temperature in the sample was stable. Strains vs. temperature curves for NR (Fig. S2), used in the calculation of entropy variations shown in Fig. 3 (main text), were measured at different constant pressures (8.7 – 173 MPa). Temperature was varied continuously by the thermostatic bath within the temperature range of ~ 285 − 330 K.

Glass-transition temperature vs. pressure data

The glass-transition temperatures ($T_g$) for NR at different pressures were measured by differential scanning calorimetry (DSC) and strain ($\varepsilon$) vs. temperature (T) curves. DSC measurement was carried out under atmospheric pressure, with heating rate of 10 K/min, from 186 K to 416 K. $\varepsilon$ vs. T curves for obtaining $T_g$ were measured in an analogous manner of $\varepsilon$ vs. T curves for $\Delta S_T$, but within a temperature range of 173–303 K; here, the $T_g$ values were obtained during heating process, when $d^3\varepsilon/dT^3 = 0$.

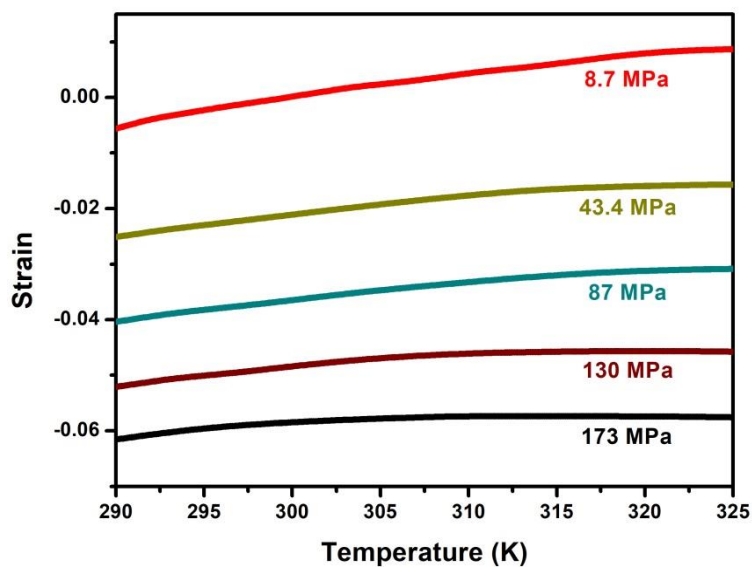

**Figure S2.** Strain *vs.* temperature curves for natural rubber at constant pressures of 8.7(2), 43.4(9), 87(2), 130(3) and 173(3) MPa measured on cooling, which were used to calculate the isothermal entropy change shown in Fig. 3 (main text).

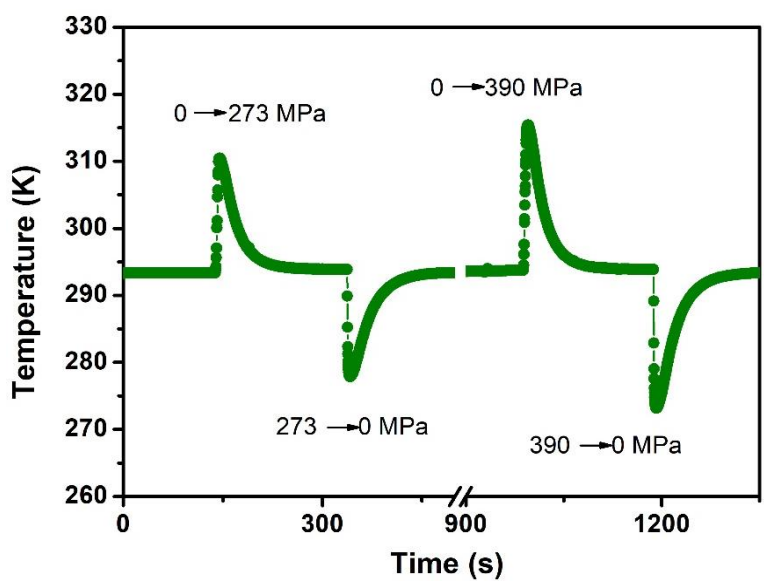

Figure S3. Temperature *vs.* time for NR at initial temperature of ~293 K; the peaks (or the valleys) are related to the adiabatic temperature change ($\Delta T_S$) when the pressures of 273(8) and 390(12) MPa are applied (or released).

**Table S1.** Fitting parameters of –$\Delta T_S$ vs. $\sigma_{max}$ curves for NR, obtained from the power law $-\Delta T(T, \sigma_{max}) = a_T \sigma_{max}^{n_T}$ (equation (2), main text).

| Temperature (K) | $a_T$ (K GPa$^{-n_T}$) | $n_T$ |
|---|---|---|
| 273 | 54(3) | 0.96(4) |
| 293 | 61(5) | 0.98(6) |
| 313 | 63(4) | 0.96(5) |

**Table S2.** Fitting parameters of –$\Delta S_T$ vs. $\sigma_{max}$ curves for NR (Fig. 4b), obtained from the power law $-\Delta S(S, \sigma_{max}) = a_S \sigma_{max}^{n_S}$ (equation (2), main text) and from the quadratic function $\Delta S(T, \sigma) = a_1(T)\sigma + a_2(T)\sigma^2$ (equation (4), main text).

| Temperature (K) | $a_S$ (kJ kg$^{-1}$ K$^{-1}$ GPa$^{-n_S}$) | $n_S$ | $a_1$ (kJ kg$^{-1}$ K$^{-1}$ GPa$^{-1}$) | $a_2$ (kJ kg$^{-1}$ K$^{-1}$ GPa$^{-2}$) |
|---|---|---|---|---|
| 295 | 0.34(1) | 0.90(1) | 0.483(2) | -0.48(2) |
| 300 | 0.29(2) | 0.88(3) | 0.447(5) | -0.54(4) |
| 305 | 0.22(3) | 0.83(6) | 0.41(1) | -0.7(1) |

*Appendix A*: **Derivation of the expression for ΔS (T,σ) from a modified Landau's theory of elasticity**

Let us regard the Helmholtz free energy per unit volume as the following series expansion:[1]

$$F(T, \varepsilon_{ij}) = F_0(T) - B\alpha(T - T_0)\varepsilon_{kk} - \frac{1}{2}B[\beta(T - T_0)^2 - 1]\varepsilon_{kk}^2 + G\left(\varepsilon_{ij} - \frac{1}{3}\varepsilon_{kk}\delta_{ij}\right)^2$$

where $F_0$ is the free energy of the unstrained samples, $\alpha$ is the thermal expansion coefficient; $\beta$ accounts for a non-linear thermal deformation of the sample; $B$ and $G$ are the bulk and shear moduli, respectively; $\delta_{ij}$ is the unit tensor. The notation above implies summation over repeated indexes (which can be *x*, *y* or *z* in Cartesian coordinates). Furthermore, $T_0$ is a reference temperature where the sample experiences no thermal deformation. The expansion above converts the components of a rank-two tensor (the strain tensor $\varepsilon_{ij}$) into a scalar.

It is possible to obtain the entropy through the derivative of the free energy with respect to temperature:

$$S = -\frac{\partial F}{\partial T} = S_0 + B\alpha\varepsilon_{kk} + B\beta(T-T_0)\varepsilon_{kk}^2 \qquad (1)$$

On the other hand, the internal stress is obtained differentiating the free energy with respect to the strain:

$$\sigma_{ij} = \frac{\partial F}{\partial \varepsilon_{ij}} = -B\alpha(T-T_0)\delta_{ij} - B\beta(T-T_0)^2\varepsilon_{kk}\delta_{ij} + B\varepsilon_{kk}\delta_{ij} + 2G\left(\varepsilon_{ij} - \frac{1}{3}\varepsilon_{kk}\delta_{ij}\right)$$

(2)

Let us now consider the case of confined compression by a uniaxial stress, and let us assume that the stress is applied along the $z$ axis. Therefore, the only non-vanishing component of the strain tensor is $\varepsilon_{zz}$. From equation (2), the component $\sigma_{zz}$ is:

$$\sigma_{zz} = -B\alpha(T-T_0) + \left\{B[1 - \beta(T-T_0)^2] + \frac{4}{3}G\right\}\varepsilon_{zz} \qquad (3)$$

Finally, combining equations (1) and (3), the entropy change can be expressed as a second-degree polynomial of the applied compressive stress:

$$\Delta S(T,\sigma) = a_1(\text{T})\sigma + a_2(\text{T})\sigma^2$$

### *Appendix B*: Satisfying the isostatic condition

The zz-component of the stress tensor above is related to the external applied stress $\sigma_{zz} = -\sigma$ (negative because the stress is compressive). The other diagonal components are:

$$\sigma_{xx} = \sigma_{yy} = -B\alpha(T-T_0) + \left\{B[1 - \beta(T-T_0)^2] - \frac{2}{3}G\right\}\varepsilon_{zz}$$

which are the components of the stress applied by the walls confining the sample and are responsible for keeping $\varepsilon_{xx} = \varepsilon_{yy} = 0$. If one considers, for the sake of simplicity, the temperature $T = T_0$, then the ratio $\sigma_{xx}/\sigma_{zz}$ becomes:

$$\frac{\sigma_{xx}}{\sigma_{zz}} = \frac{3B - 2G}{3B + 4G}$$

For natural rubber, $B = 2 \times 10^9$ Pa and $G = 3.33 \times 10^4$ Pa,[2] which results in $\sigma_{xx}/\sigma_{zz} = 0.999967$. Therefore, uniaxial compression of natural rubber is actually an isostatic compression.

**References**

1. Landau, L. D.; Lifshitz, E. M., In *Theory of Elasticity: Vol. 7 of Course of Theoretical Physics*, 2nd ed.; Pergamon Press: Oxford, 1970.

2. Tabor, D. *Polymer*. **1994**, 35, 2759– 2763.